# Objective Probability and the Mind-Body Relation

Paul Tappenden

paulpagetappenden@gmail.com

*No affiliation*

> The issue of psycho-physical parallelism is at the heart of the problem of measurement in quantum mechanics.
>
> Harvey Brown[1]

**Abstract**

Objective probability in quantum mechanics is often thought to involve a stochastic process whereby an actual future is selected from a range of possibilities. Everett's seminal idea is that all possible definite futures on the pointer basis exist as components of a macroscopic linear superposition. I demonstrate that these two conceptions of what is involved in quantum processes are linked via two alternative interpretations of the mind-body relation. This leads to a fission, rather than divergence, interpretation of Everettian theory and to a novel explanation of why a principle of indifference does not apply to self-location uncertainty for a post-measurement, pre-observation subject, just as Sebens and Carroll claim. Their *Epistemic Separability Principle* is shown to arise out of this explanation and the derivation of the Born rule for Everettian theory is thereby put on a firmer footing.

---

[1] (Brown, 1996, 189)

## 1. Two concepts of objective probability

Since the advent of quantum mechanics it has been widely thought by physicists that there may be two types of probability in the world, objective and subjective. Subjective probability is familiar as 'degree of belief' or 'credence'. It's a tool of everyday life. Objective probability is more problematic. A common term for it is 'chance', the idea of an arbitrary process selecting one possibility from a range of alternatives, but a selection guided by the alternatives' probabilities. A bridled randomness which has come to be known as stochasticity.

It can seem that Hugh Everett III's 'relative state' formulation of quantum mechanics (1957) does without a concept of objective probability. Indeed he changed the title of his thesis to *Wave Mechanics without Probability.* And some Everett theorists concur (Brown, 2011, 6; Groisman *et al.*, 2013, 696). My purpose here is to argue that there's scope for retaining a concept of objective probability in Everettian theory via an alternative to the standard stochastic interpretation of probabilistic processes. Furthermore, that alternative arises out of a startling change of perspective on the identity of observers within Everett's multiverse which helps to resolve a problematic aspect of the theory.

I shall begin with a thought experiment which suggests that there's a link between that alternative concept of objective probability and an alternative to a standard interpretation of the mind-body relation. I shall then defend the alternative mind-body relation in detail before going on to use it in an Everettian context.

The thought experiment is to take place in a setting provided by contemporary cosmology, which gives a precise meaning to the term 'parallel universes'. Space may be infinite and contain an infinite number of galaxies but there are only a trillion or two in our local region. Our *observable* universe is finite, and according to quantum mechanics any finite region can only occupy a finite number of possible observable states, so if there are an infinite number of galaxies there may be any number of regions which are exactly like our own, down to the finest observable detail (Tegmark, 2007, 104). Those regions are universes which are parallel to ours. What follows brings a change of perspective on them.

Consider a large ensemble of parallel universes in which stochastic quantum mechanics operates, that is, where a single actual outcome of a probabilistic process is understood to be stochastically selected from a range of possible outcomes. On such a view objective probabilities exist, albeit that their values can only ever be estimated via statistical methods which assume the law of large numbers.

We are to focus attention on an idealised quantum measurement where there are two possible definite outcomes on the pointer basis. A pointer on the apparatus moves left for outcome L and right for outcome R. The objective probabilities yielded by the Born rule for these outcomes are pL and pR and we can assume that those values have been statistically confirmed to a high degree of *subjective* probability.

At corresponding positions in each parallel universe we have apparatuses ready to make 'parallel counterpart' measurements. As the results come up the initial set of universes partitions into a subset where the result is L and a subset where the result is R.

Now introduce observers about to make a measurement. There are only two ways of doing this, so far as I know. The usual way is to associate an individual observer with a parallel counterpart organism in each universe. Each observer states, 'For this upcoming quantum measurement there are two possible outcomes and on statistical evidence I assign objective probabilities pL and pR to those outcomes, with subjective probability p(pL, pR)'. That statement is interpreted as being true because it refers to a stochastic process where exclusively one or the other of the outcomes will occur with probabilities pL and pR. The observer is bound to be uncertain to some degree as what the values of the probabilities are but the idea that quantum measurement involves a stochastic process implies that precise probabilities are associated with each outcome. The observers' statements are not strictly true since quantum mechanics allows for many bizarre outcomes with minute probabilities as well as the outcomes L and R, but let that pass.

A less usual way of introducing the observer is to associate a single individual with the set of parallel counterpart organisms. In that case there is just a single utterance of 'For this upcoming quantum measurement there are two possible outcomes and on statistical evidence I assign objective probabilities pL and pR to those outcomes, with subjective probability p(pL, pR)'. The parallel counterpart sonic emissions by the organisms do not each give voice to an utterance. The single utterance is voiced by the set of those sonic emissions. We have a single observer, call her Hydra. She sees a single apparatus before her which is constituted by the set of parallel counterpart apparatuses. That apparatus is going to partition into a subset where the outcome is L and a subset where the outcome is R. As a result, the parallel counterpart organisms are going to be subject to differing stimuli giving rise to cognitive differences and so the fissioning of Hydra into $Hydra_L$ who sees a pointer move left and $Hydra_R$ who sees a pointer move right.

Ted Sider has provided us with a metaphysics of transtemporal identity which is well suited to this situation (2001, 201). He introduces a concept of temporal counterparts analogous to David Lewis's modal counterparts (1968) and identifies continuant objects with momentary stages. Thus single apple resting in a fruit bowl is not the same thing from one moment to the next. Rather, at any given moment an apple bears the relation *will be* to apples which are its future counterparts and the relation *was* to apples which are its past counterparts.

So Hydra can be described as bearing the relation *will be* to each of her future temporal counterparts, $Hydra_L$ and $Hyrdra_R$, though she does not bear that relation to the pair of them. Hydra will not become two people. A modal analogy is this: suppose that you were born in Africa, then you might have been born in America (if your mother had moved there whilst pregnant) and you might have been born in Asia; but you could not have been born in America *and* in Asia.

$Hydra_L$ and $Hydra_R$, two distinct people, each bear the relation *was* to their past temporal counterpart, Hydra. The leftward pointer and the rightward pointer are sets of parallel counterparts which are future temporal counterparts of the ready pointer. True, Sider's stage theory has the odd consequence that many people have worked on writing these very words but it's arguably not impossibly odd since they are all people who I, now, was. Likewise, there would have been many apples resting in the fruit bowl overnight, though only one apple and one bowl at any given moment.

In the spirit of Donald Davidson's 'radical interpretation' (1973) Hydra can be interpreted as speaking truly when she makes her single utterance of 'For this upcoming quantum measurement there are two possible outcomes and on statistical evidence I assign objective probabilities pL and pR to those outcomes, with subjective probability p(pL, pR)'. What she refers to is an apparatus which will fission into subset apparatuses where L and R occur. What she refers to as possibilities are multiple future actualities which are *causally connected* with her perceived environment, and what she refers to as the objective probabilities of those possibilities are her estimation of the measures of the L and R subsets of her apparatus relative to the set which is the apparatus in the ready state. There will be more on causality in Hydra's environment in the next section.

What this suggests is that a concept of a *non-modal* objective probability is intelligible; a concept of objective probabilities which attach to a range of actualities rather than of possibilities. This may seem to be flirting with absurdity. Before even beginning to seriously entertain the idea it must be established that the alternative 'unitary interpretation of mind' is itself intelligible, which I shall attempt to do in the next section. It is a radical proposal which requires careful scrutiny, but it has long been thought that making sense of a reality underpinning quantum phenomena will require a radical conceptual shift.

**2. The unitary interpretation of mind**

The idea that 'a plurality of worlds' exists which contains worlds parallel to ours has been around for a long time and it has always seemed natural to think of those parallel worlds as far off in the distance but if we adopt Hydra's perspective they are all right here. In some sense our perceived environment must be a sort of 'superposition' of parallel universes if such exist.

Gottfried Leibniz put these words into the mouth of an interlocutor in a dialogue:

> what is to prevent us from saying that these two persons who are at the same time in these two similar but inexpressibly distant spheres, are not one and the same person? Yet that would be a manifest absurdity.
>
> (Leibniz, 1704, Bk.II, Ch.xxvii, 245)

This expresses exactly the thought in the Hydra scenario. More recently the idea has been discussed in (Zuboff, 1974, 374; 1991, 41-2; Bostrom, 2006, 186-8). A much fuller development is to be found in (Tappenden, 2011a, sections 2, 4 and 5 ) which I shall summarise here.

First of all, it can indeed seem 'manifestly absurd' that parallel counterpart organisms vastly separated in space could be multiple instances of a single mind if it is thought that there must be some sort of causal connection between them. But all that radical interpretation requires is that the organisms and the environments with which they interact should be isomorphic. With that in mind we can approach interpreting Hydra's speech and behaviour.

Hydra says 'I see a single apparatus before me which has a mass of one kilogram'. For this to be interpreted as true she cannot be referring to the aggregate of the parallel counterpart apparatuses since that has a much greater mass, but another type of collective is available, the set of the apparatuses. Usually sets are thought to be abstract but that is not a requirement. Willard Van Orman Quine suggested that some sets could be regarded as concrete when he wrote:

> none of the utility of class theory is impaired by counting an individual, its unit class, the class of that unit class, and so on, as one and the same thing.
>
> (1969, 31)

Quine's proposal violates the Axiom of Foundation and Lewis's 'Main Thesis' that 'the parts of a class are all and only its subclasses' (1991, 7), and has not in fact been made use of in set theory but there is no argument that I know of which dismisses it as unintelligible. So it appears possible that Hydra can be understood to be referring to a single apparatus in her perceived environment which is constituted by a set of parallel counterpart apparatuses each of which is its own sole element. In that case we can understand Hydra to perceive her apparatus as having the mass which all its elements have in common, i.e. one kilogram.

To emphasise this point, imagine that into the ensemble of parallel universes which Hydra's mind spans we were able to introduce parallel counterpart black boxes with causally isolated interiors so that they could contain anisomorphic contents. The contents of the box in Hydra's environment would be a set with anisomorphic elements. If Hydra were to open the box each of the parallel counterpart organisms which are elements of her body would move in concert and, on receiving different stimuli from the box's contents, cognitive change would be induced so causing Hydra to fission. 'Oh, it's a duck!' Hydra$_{DUCK}$ would exclaim, and 'Oh, it's a rabbit!' Hydra$_{RABBIT}$. This makes it clear that an object in Hydra's environment can only be perceived by her to have definite physical properties if all the elements of that object have those properties in common.

Parallel observable universes in an infinite space are separated by vast spatial distances and simultaneity is relative so in what sense does Hydra's reference to places and times relate to space and time? She says, 'My laboratory is about one mile NNE of the Big Ben clock tower which is showing four o'clock'. Charitable interpretation allows us to understand that she's referring to a clock tower which is a set of parallel counterpart clock towers each of which indicates corresponding places and times in the parallel universes which her mind spans. The times and places to which Hydra refers in her perceived environment are sets of parallel counterpart times and places.

A worry may remain. We commonly understand a person's action to be caused by beliefs and desires. Hydra believes that she has a quantum measurement apparatus before her and desires to operate it, which is why she extends an index finger to press the button. When she acts, all the doppelgangers which are elements of her body move in concert to extend fingers towards buttons. Shouldn't each of those actions be explained by a local mental causation, implying the standard 'plural' interpretation of mind which would locate an observer in each parallel universe?

That's an option, but it's not necessary. Just as Hydra's utterance can be understood to be expressed by a set of parallel counterpart sonic emissions, so her intentional action can be understood to be expressed by a set of parallel counterpart bodily motions. If each bodily motion is taken to be locally caused by neural activity the question then remains as to how that neural activity is related to mentality. Does each brain instance a distinct mind or are all isomorphic brains instances of a single mind just as they are instances of a single physical form? It is the physical form of that neural activity which determines its causal powers. I should stress that no very fundamental distinction between minds and non-minds is being made here. If minds are supposed to arise out of the structural properties of brains then the idea is simply that numerically distinct brains with isomorphic structure instance a single mind.

Now consider causality in Hydra's environment. When she presses the button on her apparatus it causes the quantum measurement to be made. If causality is thought of as a mysterious relation of natural necessitation between the button and the rest of the apparatus then causation in Hydra's environment is the set of individual causal relations. If causation is thought of as constant conjunction then, since the universes are parallel, a constant conjunction of events in each universe will be a constant conjunction of sets of events in Hydra's environment. When constant conjunction ceases for a set of parallel universes they 'diverge', which is to say that they become anisomorphic.

From now on I shall assume that the unitary interpretation of mind is an intelligible alternative to the standard plural interpretation which holds that minds may be qualitatively identical and numerically distinct. In that case, what the thought experiment with Hydra shows is that it's intelligible for an observer to believe that when they conduct a quantum measurement with multiple outcomes they and their measuring device will fission into different branches of reality, each outcome occurring in a different branch. The modal interpretation of objective probability, which assumes that what happens in a quantum process is that one of a range of possibilities becomes actual, is revealed as a conjecture for which there is an alternative: the non-modal interpretation which assumes that what happens is fission, objective probabilities attaching to actualities issuing from a common cause.

If Hydra comes to understand this, she's free to drop the assumption that quantum evolution is stochastic. Even if she happens to inhabit one dendritic multiverse rather than an ensemble of cosmological parallel universes she is free to agree with what Everett notoriously wrote:

> The whole issue of the translation from 'possible' to 'actual' is taken care of in the theory in a very simple way—there is no such transition, nor is such a transition necessary for the theory to be in accord with our experience. From the viewpoint of the theory *all* elements of a superposition (all 'branches') are 'actual', none any more 'real' than the rest.
>
> (1957, 459, his emphasis)[2]

The theory to which Everett refers is his 'relative state' formulation of quantum mechanics. In its modern version, in which the process of decoherence effectively defines the pointer basis, the idealised quantum measurement with outcomes L and R is understood as follows (Wallace, 2012, 74-102). The measurement process, rather than being stochastic, involves the evolution of the measuring device into a linear superposition of apparatuses for which the outcomes L and R occur. Each of these components of the superposition a 'branch' of the multiverse and each branch has a quantum amplitude. It is the squared modului of those amplitudes, as proportions of their total for all the branches, which the Born rule can be understood to interpret as an objective probability.[3]

---

[2] Note that Everett is not here using the term 'actual' in an indexical sense, as in Lewis's 'modal realism' (1986) and adopted in (Wilson, 2013), of which more later.

Hydra in the cosmological context fissions into a branch of reality where L occurs and a branch of reality where R occurs and yet she said, 'For this upcoming quantum measurement there are two possible outcomes and on statistical evidence I assign objective probabilities pL and pR to those outcomes, with subjective probability p(pL, pR)'. If she is to be charitably interpreted as speaking truly her term 'possible outcomes' has to be taken to refer to actual outcomes in her branching future. But to translate 'possible' as 'actual' would seem to undermine our modal discourse which would surely be an unacceptable consequence. More discussion is needed.

## 3. Modal and non-modal objective probabilities

We need to clarify Hydra's use of the term 'possible' in that cosmological ensemble of parallel universes where stochastic quantum mechanics hypothetically operates. So, rather than focussing on a quantum measurement device in each universe we focus on a massive ratchet wheel which, like a roulette wheel, has numerals from '0' to '36' marked evenly around the periphery and a pointer beside which one or another number always comes to rest. With an average push the wheel takes a few tens of seconds to stop but once set in motion a linked apparatus consisting of a motion detector and computer is able to reliably predict which number will come to rest beside the pointer. Of course there will be quantum effects so the apparatus cannot be perfectly reliable but suppose that the unreliability is negligible. Assuming the law of large numbers, when a large ensemble of parallel counterpart ratchet wheels are set in motion the ensemble partitions into subsets where different numerals come to rest beside the pointer but for all but one of the numerals the measure of the subset of universes where that numeral comes to rest is very small. As before, set aside these highly improbable stochastic outcomes; the partitioning is hugely dominated by a subset of universes which has a measure of almost unity.

---

[3] This is a simplification since any given outcome branch will consist of very many microscopically different 'sub-branches' but that will not be important in the following argument.

Hydra is now cast in the role of a punter who knows nothing of quantum mechanics and who is invited to gamble with the ratchet-wheel as she might with an ordinary roulette wheel. Bets must be placed within a few seconds of the wheel being set in motion, at which time Hydra hasn't the slightest idea which number will come up, the readout on the prediction apparatus not being accessible. As the parallel counterpart wheels which are elements of Hydra's wheel are set in motion she says, 'For this current spin of the wheel there are 37 possible outcomes and the probabilities for each are 1/37'.

Recall that the charitable interpretation required by the unitary interpretation of mind in the context of a quantum measurement obliges us to translate Hydra's term 'possible outcome' as 'actual outcome' but if the same translation is used here Hydra is clearly mistaken. Once her wheel is set in motion there is, for all practical purposes, just one actual outcome whose objective probability is unity. That is the *quantum-mechanical* objective probability for the outcome which would be predicted by the computer linked to a motion detector. The ratchet wheel behaves *as if* it were a classical device but its behaviour is strictly quantum-mechanical. In other words it is a 'quasi-classical' device.

To speak of possible outcomes and their probabilities is a very common and natural way of speaking in roulette wheel contexts. What has happened is that Hydra has assigned *subjective* probabilities of 1/37 (hereafter 'credences') on the basis of what are *imagined* to be objective probabilities, often called chances, attaching to entities referred to as possibilities. The true objective probabilities, which are quantum-mechanical, apply to the actual branching of the ratchet wheel, which is hugely dominated by a single branch whose measure is almost unity. That is to say, the ratchet wheel in Hydra's environment branches into subsets and one of those subsets has a measure close to unity relative the set of ratchet wheels originally sent spinning, always assuming the law of large numbers.

We have Hydra facing a ratchet wheel which has just been given a push and which is a set of parallel counterpart ratchet wheels. For all practical purposes Hydra does not fission when her ratchet wheel comes to rest. More accurately, she does fission but all but one of the branches into which she fissions is a subset of parallel universes with negligible measure on the entire set. There are not 37 branches of equal measure where each of the outcomes from 0 to 37 occurs. So when Hydra says, 'For this current spin of the wheel there are 37 possible outcomes and the probabilities for each are 1/37' she is clearly not assigning objective probabilities as in the case of the quantum measurement.

The ratchet wheel shows the way to preserve ordinary modal discourse whilst proposing that objective probabilities, which only arise in quantum mechanics, should be assigned to actualities, not possibilities. If asked why she has a credence of 1/37 that the number 7 will come up Hydra may reply that the reason is that the (objective) 'chance' of that number coming up is 1/37 but there's not an objective probability of 1/37 that the spinning ratchet wheel will come to rest with the number 7 by the pointer. That's a fiction. There's an objective probability of almost unity that a particular number will come up, which may or may not be 7.

So there are two distinct domains of discourse about objective probabilities. In one domain they are assigned to imaginary 'possibilities' which are projected out into the world as supposed warrants for credences, as in the case of Hydra's attribution of outcome probabilities for the ratchet wheel, which she calls chances. That is the modal application of the term 'objective probability'. What the existential status of those imagined possibilities is is a matter for modal theory. If they are understood to be real then they must be considered non-actual since only that part of reality described by quantum mechanics is what is being taken to be actual.

But when Hydra assigned a probability of $p_R$ to the pointer pointing right she was assigning an objective probability to an actual outcome, a future branch of her environment. That is the non-modal and quantum-mechanical use of the term 'objective probability' which applies to actualities, not possibilities.

The use of the unitary interpretation of mind in the context of a large ensemble of stochastic parallel universes shows that the concept of objective probabilities as values attaching to actual future branches of reality is intelligible. We are thus free to abandon the stochastic interpretation of objective probability and replace it with the dendritic interpretation. However, some lacunae in the picture need filling in.

**4. Uncertainty lost and regained**

We make decisions about future-directed action on the basis of credences, which are subjective values associated with alternative possibilities. That is particularly clear for quasi-classical games of chance where the alternative possibilities are assigned precise probabilities often commonly thought of as objective chances, as in the case of the ratchet wheel. In betting scenarios, ever the life blood of probability theory, it is generally accepted that the credences attaching to alternative possibilities should equal the presumed objective probabilities (chances) of those alternatives, what Lewis dubbed the Principal Principle (1980, 266). On placing a wager a punter is said to be *uncertain* which of the alternative possibilities will become actual even if s/he feels certain about what the chances are for each. To be certain about an outcome is to assign it a credence of unity.

But uncertainty apparently disappears when probabilistic processes are understood to be dendritic rather than stochastic since the concept of alternative possible outcomes is replaced by that of co-actual outcomes. And this can seem to undermine any reason to place wagers for a person believing the dendritic interpretation of probability. To illustrate, imagine Hydra again, now in the context of an Everettian branching multiverse rather than a cosmological ensemble of parallel universes. Presume that she's aware of her situation and is invited to place a wager at given odds on that idealised quantum measurement where there are two outcomes, L and R with objective probabilities $p_L$ and $p_R$. If she believes that the apparatus and she herself will fission there are no possibilities in the offing with which she can associate credences less than unity. She is simply certain that fission will take place and the distribution of losses and gains across a range of actual outcomes appears to be quite different from their distribution across a range of alternative possibilities. It is not at all obvious that Hydra has any reason to place a bet.

There are two ways in which theorists have attempted retrieve uncertainty in this sort of situation, both of which involve the concept of self-location uncertainty. One of those ways is via a 'divergence' interpretation of Everettian theory. The basic idea is that branching is to be thought of as like a bundle of fibres diverging into tresses. In a quantum measurement situation an observer is one of a multitude of exact copies, each in what David Deutsch called 'identical universes' (1985, 20). As the measurement takes place the universes diverge into subsets where different outcomes occur but prior to the measurement each observer has self-location uncertainty as to which sort of universe they inhabit. Recent variants on the theme have been proposed in (Saunders and Wallace, 2008; Saunders, 2010; Wilson, 2013). Clearly this is incompatible with the unitary interpretation of mind since the mind of a subject will span identical universes rather than inhabiting them individually so I shall not discuss divergence further.

A second way of introducing self-location uncertainty to Everettian theory was first noted by Lev Vaidman (1998, 253). Suppose that Hydra is knowingly about to make the idealised branching measurement as before but she's blindfolded so that she can't see which way the pointer moves. Vaidman's idea is that she fissions into $Hydra_L$ on the L branch and $Hydra_R$ on the R branch and that they are then each uncertain as to which branch they're on. This idea needs some modification in the light of the unitary interpretation of mind because Hydra will not fission until some cognitive difference arises and I shall come to discussing that later. Setting that thought aside, Vaidman's idea confronts two problems.

The first problem is that post-measurement, pre-observation self-location uncertainty doesn't seem to help in providing a reason for Hydra to place a wager before the measurement since it comes too late, if it comes at all. This problem is addressed in (Tappenden, 2011b) and here is a brief summary of the proposed solution. Hydra is told that she must lay a stake on one of the outcomes before the measurement in order for there to be a payout on that branch and she knows that it's possible for her to be uncertain about which branch she's on after the measurement. In that post-measurement situation, since she would know the probabilities and offered payout, she could decide whether she wanted to lay a stake or not but of course it's too late. However, Hydra knows before the measurement that if she were in a state of Vaidmanian uncertainty after the measurement and decided that it was worth laying a stake she would regret not having laid the stake beforehand. Thus Hydra has a good reason to lay a stake before the measurement if she believes that in the Vaidamanian situation she would judge the bet worth taking. Avoiding possible future regret provides a good reason to act. However, this does not imply that future-directed probabilistic decisions will always be unaffected by a shift from a standard stochastic view of quantum mechanics to a fission interpretation of Everett's theory. Scenarios such as that suggested by Huw Price (2010, Section 6) may give rise to bizarre conundrums and there is the notorious phenomenon of quantum Russian roulette waiting in the wings (Tappenden, 2004, 158).

Another problem for post-measurement, pre-observation, self-location uncertainty is this. If blindfolded Hydra makes her measurement and fissions into Hydra$_L$ and Hydra$_R$, each has to decide what credence to accord to the possibilities of being on one or the other branch. For self-location uncertainty a principle of indifference seems compelling and has been argued for in detail by Adam Elga (2004). That implies that Hydra$_L$ and Hydra$_R$ each have no reason to suppose that she is on one branch rather than the other. In which case they should each assign a credence of ½ to being on one or the other branch *irrespective of the values of pL and pR*. That is of course a disastrous result since what is supposed to be the objective probability of outcomes L and R, the relative squared moduli of amplitude for their branches, is irrelevant to Hydra$_L$'s and Hydra$_R$'s credence assignments.

This problem has the potential to scupper the Everettian project in one fell swoop. Within a branching quantum multiverse a post-measurement, pre-observation subject is necessarily ignorant as to which type of outcome branch s/he's on. And Everett's idea can only make sense if such a subject assigns credences equal to the Born values for the outcome branches. Simply assuming that, and thus dismissing the principle of indifference by *fiat*, is what I have called the Born-Vaidman rule (2011b, Section 2).

Sebens and Carroll have attempted to do better, arguing that the principle of indifference does not apply to self-location uncertainty in the post-measurement, pre-observation context because of what they call an Epistemic Separability Principle (ESP). And they go on to claim that from the ESP it can be shown that the credences Hydra$_L$ and Hydra$_R$ ought to assign to being on the L or R branch should be determined by the Born values of those branches. I shall now discuss Seben's and Carroll's argument in detail before going on to argue that their ESP is entailed by the unitary interpretation of mind.

## 5. The Epistemic Separability Principle

The framework which Sebens and Carroll use for their discussion is an adaptation of a thought experiment used in (Elga, 2000). They make it clear that this is an idealised scenario but it suffices to present the problem and their solution to it (*ibid.*, 13-14).

*Once-or-Twice*

> Alice's particle (*a*) and Bob's particle (*b*) are both initially prepared in the *x*-spin up eigenstate. Alice's device measures the *z*-spin of her particle first. Then, Bob's device, which is connected to Alice's, measures *z*-spin of particle *b* only if particle *a* was measured to have *z*-spin up. By $t_1$, the setup is prepared; by $t_2$, Alice's particle has been measured but Bob's has not; by $t_3$, both particles have been measured. Bob has been watching as the results of the experiments are recorded. Up through $t_3$, Alice has not looked at the measuring devices and is unaware of the results. By $t_4$, Alice has looked at her device and seen the result of the measurement of particle *a*, although she remains ignorant about the *z*-spin of particle *b*. The branching structure of this scenario is shown [below].

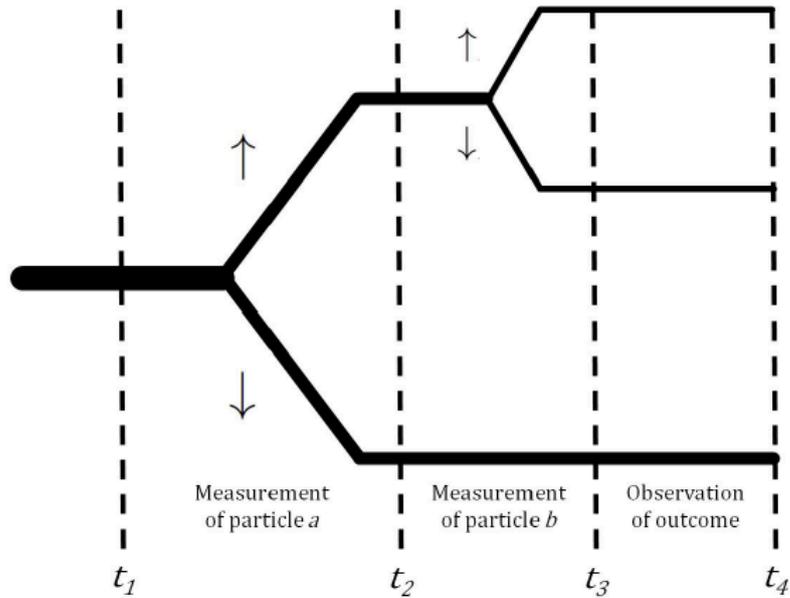

(*ibid.*, 8)

Sebens' and Carroll's aim is to make the fission interpretation of Everettian theory intelligible so they interpret Once-or-Twice in the following way:

> There are two copies of Alice at t2 in Once-or-Twice. Each copy can reasonably wonder which one she is. Thus even if she (incredibly) knows the universal wave function exactly, Alice still has something to be uncertain of. She isn't uncertain about the way the universe is; by supposition, she knows the wave function and this gives a complete specification of the state of the universe. Alice is uncertain about where she is in the quantum multiverse … She doesn't know if she's in the branch of the wave function in which the detector displays up or the one in which it shows down.

(*ibid.*, 11)

If the principle of indifference is to guide the downstream Alices' credences as to the result of the measurement of particle *a* it looks as though the two Alices at $t_2$ should assign credences of 1/2 to each possibility and the three Alices at $t_3$ should assign credences of 2/3 and 1/3 (*ibid.*, 14). And yet all that has happened between $t_2$ and $t_3$ is that Bob has made the measurement on particle *b* which has no effect on the quantum amplitudes of the two downstream branches for particle *a*. Sebens and Carroll respond to this as follows:

> If indifference is right, there's a strange switch in the probabilities between $t_2$ and $t_3$. Is there any reason to think this undermines the branch-counting strategy advocated by indifference? Wallace has argued that such a switch violates a constraint he calls 'diachronic consistency'. In Appendix A, we argue that this is not the right diagnosis of the problem with the switch in credences. This kind of inconsistency is a common result of indifference and not something that should be taken to refute the principle. Still, we agree that there's something wrong with the probability switch.
>
> (*ibid.,*15)

In attempting to put the 'something wrong' right, Sebens and Carroll aim to escape this Vaidmanian *impasse* and in preparing the ground to introduce ESP they write:

> Between $t_2$ and $t_3$ what happens? Particle b is measured and Bob takes note of the result. Nothing happens to Alice, particle a, or Alice's device. If nothing about Alice or her detector changes, why should her degree of belief that she bears a certain relation to the detector change? … Why should her

> probability for being in different subsystems … of the (Alice + Dectector) system change when nothing about that system changes and she knows that she is somewhere in that system? It shouldn't.
>
> (*ibid.,* 15)

To a critical eye this looks odd. In what sense does 'nothing happen' to Alice, particle *a* and her device between $t_2$ and $t_3$? She transits from having two copies to having three! However, there is a sense in which nothing happens : there is no cognitive change in Alice and no change in the *relation* between each copy of Alice and each copy of particle *a* and her device. All that has happened is that there has been a change in the environment exterior to the Alice-particle(*a*)-device subsystem, namely due to Bob's measurement of particle *b*.

Sebens' and Carroll's ESP captures that thought to suggest that the downstream Alices should not change their credences as to which branch of the Alice-particle(*a*)-device subsystem they are in:

> ESP: Suppose that universe *U* contains within it a set of subsystems, *S*; such that every agent in an internally qualitatively identical state to agent *A* is located in some subsystem that is an element of *S*. The probability that *A* ought to assign to being located in a particular subsystem, $X \in S$, given that they are in *U*, is identical in any possible universe that also contains subsystems *S* in the same exact states (and does not contain any copies of the agent in an internally qualitatively identical state that are not located in *S*).
>
> (*ibid.*, 16)

Nonetheless a critic may well insist that Sebens and Carroll have here simply stipulated their way out of the *impasse* by providing a principle specifically designed to do what they want, witness the 'ought', which extends to a 'Strong ESP' to preserve the principle of indifference for within-branch contexts (*ibid.,* 24).

What the fission interpretation of Everettian theory needs is a more fundamental reason to accept ESP and reject indifference. The unitary interpretation of mind can provide just that. I shall now demonstrate why it entails ESP and why it's not important that the principle of indifference is thereby lost for within-branch contexts.

**6. Alice in wonderland**

Here is the Once-or-Twice setup again:

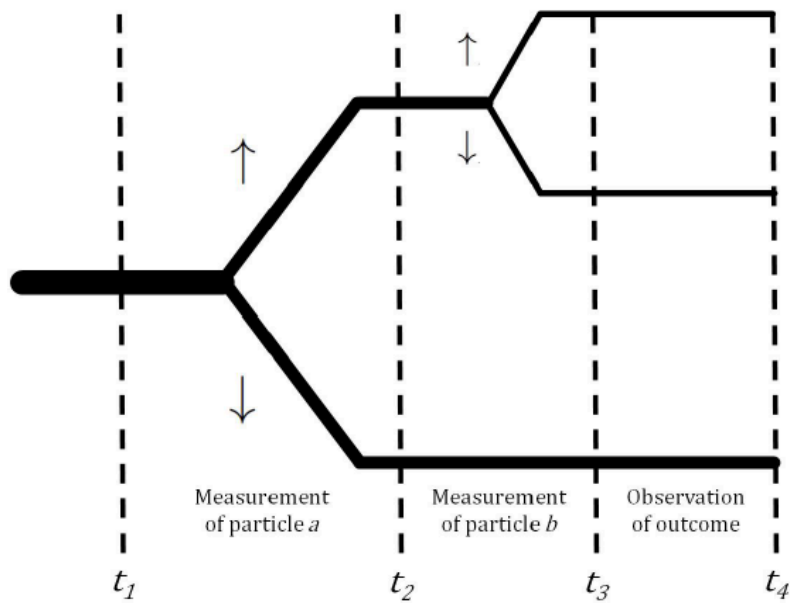

And here's the analysis applying the unitary interpretation of mind. Between $t_1$ and $t_2$ the measuring device for particle $a$ fissions into two devices, one showing the result *up* and the other showing the result *down*, each having the same amplitude. As a result Alice's body fissions also, but her *mind* does not since she has not become aware of the outcome of the measurement of particle $a$, nor has any consequence of that measurement had any sensory effect for her.

There's an alternative way to put this, responding to the point made by Adrian Kent that nothing in the formalism forces us to speak of the fissioning of the measurement device and Alice's body (2015, 214). We can simply say that the measurement device evolves into a superposition of showing *up* and *down*, and that Alice's body, on becoming entangled with that superposition, also evolves into a superposition on the pointer basis. However, Alice remains a single subject because there is no sensory or cognitive difference between the components of her bodily superposition.

Taking the analysis of Hydra in parallel universes as a guide, at $t_2$ Alice's body would a doubleton set with one element on the *a-up* branch and one element on the *a-down* branch and the measuring device in Alice's environment would also be a set of two measuring devices, each with the same quantum amplitude. However, for Alice in a branching multiverse the set-theoretic analysis is not required. The reason is that the concept of linear superposition introduces a new type of part-whole relationship to physics. The components of a superposition are neither spatial parts nor temporal parts but they can be considered to be 'superpositional' parts[4].

What is more, the relationship between a superposition and its component parts is such that the superposition only has definite physical properties if all its parts share that property, exactly as was required for the set-theoretic analysis of Hydra in the context of multiple parallel universes. Of course, cosmological parallel universes may still exist in an Everettian context and so the set-theoretic analysis may still have a place, but since we are now considering a single *causally isolated and spatially finite* "observable" universe, that thought can be set aside. The upshot is that at $t_2$ the measuring device in Alice's environment is in an equal-amplitude *macroscopic linear superposition* of showing *up* and *down*.

Between $t_2$ and $t_3$ the *b*-device on the *a-up* branch fissions causing the superpositional component of Alice's body on the *a-up* branch to fission so Alice's body now has three components, two on the *a-up* branch and one on the *a-down* branch. However, the operation of the *b*-device has no effect on the quantum amplitudes of the *a-up* and *a-down* branches. Then Alice observers the *a*-device and that causes a cognitive difference to arise between the component of her body on the *a-down* branch and the two components of her body on the *a-up* branch. As a consequence, Alice fissions into Alice$_{UP}$ who observes *a-up* and Alice$_{DOWN}$ who observes *a-down*. For Alice$_{UP}$ the *b*-device is still in linear superposition.

---

[4] In (Tappenden, 2000, 105) I refer to those parts as *superslices*, being parts of *super*positions which bear some resemblance to so-called time*slices*.

It's now clear why the unitary mind analysis of Sebens' and Carroll's Once-or-Twice entails their ESP. At $t_3$ Alice's environment includes macroscopic linear superpositions of both the *a*-pointer and the *b*-pointer but the amplitudes of the *a*-pointer superposition are in no way affected by the amplitudes of the *b*-pointer superposition. Between $t_3$ and $t_4$ Alice's mind fissions because she observes the *a*-pointer but she does not observe the *b*-pointer so that remains in superposition. Alice fissions into Alice$_{UP}$ for whom the *a*-pointer indicates *up* (and for whom the *b*-pointer is in superposition) and Alice$_{DOWN}$ for whom the *b*-pointer is not in superposition as it remains in the ready state. Alice$_{UP}$'s body thus has two superpositional components.

So, what we have seen is that the unitary interpretation of mind, applied in the context of a cosmological ensemble of spatially separated parallel universes for which a stochastic conception of objective probability is assumed, demonstrates that an alternative dendritic conception of probability is intelligible. And that is so despite the strong intuition that the very meaning of the term 'objective probability' involves stochasticity. Furthermore, the unitary interpretation of mind entails Sebens' and Carroll's ESP from which they derive the Born rule for Everettian theory.

However, recall that it was claimed that to have reason to place a wager prior to a measurement Alice would need to be able to appeal to post-measurement, pre-observation *uncertainty* and from the unitary interpretation of mind it follows that Alice post-measurement, pre-observation is not in a state of uncertainty. But uncertainty is easily recovered. At $t_2$ let a bell ring on the *a-up* branch and a whistle blow on the *a-down* branch without Alice knowing which sound goes with which branch. She will fission and although the resulting Alice$_{UP}$ and Alice$_{DOWN}$ will not be cognitively identical they will still be able to wonder which branch each is on.

Notice that the 'possiblities' of self-location are fictional in the sense that Alice$_{UP}$ is actually on the *up* branch and Alice$_{DOWN}$ is actually on the down branch. The right question for each to be asking is, 'What is the probability of the branch I'm on relative to the measurement event?'. That is to take a branch's probability as being a relation between physical properties which can be thought of as a novel 'superpositional' form of extension.

Having introduced the ESP to block the use of the principle of indifference in which-branch contexts Sebens and Carroll go on to introduce a 'strengthened' ESP to *preserve* indifference for within-branch contexts. On the face of it, this looks problematic for the unitary interpretation of mind which excludes the use of the principle of indifference in any context which presumes the existence of qualitatively identical and numerically distinct minds. However the apparent conflict has no consequence if the unitary interpretation of mind yields the same credences for within-branch contexts as does the principle of indifference, which I shall argue is the case. The problem can be set up by adapting an example introduced in (Elga, 2004) and used by Sebens and Carroll (*op cit.,* 13).

**7. Dr. Evil in wonderland**

Dr. Evil is plotting the destruction of Earth from his lunar battle station when he receives an unwelcome message. Back on Earth some pesky philosophers have created two copies of the entirety of his battle station, perfectly replicating every piece of furniture, every weapon, and every piece of food, even replicating the stale moon air and somehow the weaker gravitational field. They went so far that at time *t* they created two copies of Dr. Evil's body.

According to the standard plural interpretation of mind there are three people involved here, Dr. Evil on the moon and two people with qualitatively identical minds to his on Earth. From the principle of indifference it follows that Dr. Evil should assign a credence of two thirds to seeing a terrestrial landscape when he opens the battle station door because he could just as well be any one of the three people.

On the unitary interpretation there is one person, Dr. Evil, whose body has three isomorphic elements. His battle station has three isomorphic elements too but the environment beyond is different, it is what might be called a classical superposition, like the contents of Hydra's black box in cosmological parallel universes. The external environment has one element which is lunar and two elements which are terrestrial. If Dr. Evil believes the unitary interpretation then he believes that on opening the door he will fission into Lunar Evil seeing a lunar landscape and Earthly Evil seeing a terrestrial landscape and that Lunar Evil will have a body with one element and Earthly Evil a body with two elements. From this perspective, what credence should Dr. Evil assign to seeing a terrestrial landscape when he opens the door?

He can reason as follows. Suppose his mind spanned a large ensemble of cosmological stochastic parallel universes and suppose that a stochastic process with two possible outcomes, *moon* and *earth*, were to take place outside the battle station where the probability p*moon*=1/3 and p*earth*=2/3. In that case, assuming the law of large numbers, he should assign a credence of 2/3 to seeing the outcome *earth* on opening the door since that is the objective probability for that outcome.

So Dr. Evil can conclude that if his environment is a classical superposition with a finite number of elements then the proportions of those elements corresponding to each component of that superposition should be treated *as if* they were objective probabilities and so guide his credence assignments to the seeing of different outcomes on observing the superposition. To be sure, the creation of the duplicate battle stations has not been a stochastic process but what the idea of a large, finite, stochastic ensemble of universes shows for the unitary interpretation of mind is that the proportions of elements in a finite classical superposition can be regarded *as if* they were objective probabilities.

That suggests that Dr.Evil should treat the unobserved classical superposition of the lunar and terrestrial environments with one and two elements respectively *as if* it were a linear superposition with those same

Born values, in which case he should assign a credence of 2/3 to seeing a terrestrial environment on opening the door, just like the three Dr. Evil doppelgangers of the plural interpretation of mind.

**8. Parting Lines**

What is being proposed is a radical change of perspective. Macroscopic linear superpositions can easily exist in the environment of an observer, only requiring that s/he be perceptually isolated from quantum measurements and similar decoherence phenomena in her past light cone. A person's brain can be a macroscopic linear superposition so long as there are no sensory or cognitive differences between its components (conscious or unconscious). Probabilistic processes are to be thought of as dendritic rather than stochastic, implying no distinction between possibility and actuality *within an Everettian multiverse*, just as Everett emphasised in his famous footnote. Objective probability inhabits the actual world, not the realm of possibilia.

That's why, when blindfolded Alice hears a bell ring and asks herself what credence to give to being in a branch with relative probability pR or pL she takes what Vaidman calls the measure or existence of the branches into account (1998, Section 9; Groisman, Hallakoun and Vaidman, 2013). Vaidman rejects stochastic probability but there is no reason for him to reject dendritic probability, in which case objective probability just *is* a relation between the measures of existence of branches. And the objective probability of her branch relative to the measurement event is what, by any reasonable standards, guides Alice's credence that she's on it.

What emerges is that Everettian theory is not so much an interpretation of quantum mechanics as an interpretation of the very concept of objective probability itself. Decoherence is a discovery independent of Everett's theory and perfectly compatible with a stochastic metaphysics as is well demonstrated by Murray Gell-Mann and James Hartle (2011). So decoherence does not bring some sort of Everettian influence to quantum mechanics. It's the other way round; decoherence simply provides an approximate pointer basis which picks out the dendritic structure which Everett had in mind.

This is all very counterintuitive. It seems compelling that if we are about to place a bet on a quantum measurement with given probabilities for outcomes we must be uncertain about what will happen. But that turns out not to be so, the possibility of being able to make post-measurement, pre-observation credence assignments can do all the work of giving a reason to place the bet before the measurement.

A final point worth making is this. The unitary interpretation of mind provides a unique defence of semantic internalism against the very influential challenge of Twin Earth thought experiments, as is argued in (Tappenden, 2011a). And a currently very promising theory of mind, Prediction Error Minimization, apparently requires semantic internalism (Hohwy, 2016, 24, note 8 ). If semantic internalism is correct then the meaning of the term 'probabilistic process' cannot be determined by the constitution of the external world. In which case probabilistic processes cannot be assumed to be stochastic just because that is taken to be what we commonly mean by the term which refers to them. The idea that if objective probability exists it must be stochastic is no more than a hypothesis about the world and we have seen that an alternative hypothesis is available, that of dendritic probability.

The unitary interpretation of mind has many other implications. It brings a novel perspective to the Boltzmann-brain and measure problems in cosmology and to philosophical problems to do with the concepts of unconscious 'zombies', personal teleportation and mind uploading.

## Acknowledgements

Special thanks to Harvey Brown, Charles Sebens and Alastair Wilson for searching comments which lead to important changes in earlier drafts. Thanks also to Douglas Campbell, Thomas Forster, and Douglas Porpora for some very useful remarks, and to Simon Saunders for many discussions and for finding the Leibniz quote.

This research did not receive any specific grant from funding agencies in the public, commercial or not-for-profit sectors.